\documentclass[a4paper,11pt]{article}
\usepackage{pos}
\usepackage{amsmath}
\usepackage{dsfont}
\usepackage{hyperref}
\usepackage{feynmp}
\usepackage{float}
\usepackage{tikz-feynman}
\tikzfeynmanset{compat=1.1.0}
\usepackage{slashed}
\usepackage{float}
\usepackage{appendix}
\usepackage{dirtytalk}
\usepackage{hyperref}
\title{ Re-evaluating Jet Reconstruction Techniques for New Higgs Boson Searches}

\author[a]{A. Chakraborty}
\author[b]{S. Dasmahapatra}
\author[e]{Henry A. Day-Hall}
\author[c]{Billy Ford}
\author*[c]{S. Jain}
\author[c,f]{ S. Moretti}
\author[d]{ E. Olaiya}
\author[d]{ C.H. Shepherd-Themistocleous}

\affiliation[a]{Department of Physics, School of Engineering and Sciences, SRM University AP,\\
  Amaravati, Mangalagiri 522240, India}

\affiliation[b]{School of Electronics and Computer Science, University of Southampton,\\
Southampton, SO17 1BJ, U.K}

\affiliation[c]{School of Physics and Astronomy, University of Southampton,\\
  Southampton, SO17 1BJ, United Kingdom}

\affiliation[d]{Particle Physics Department, Rutherford Appleton Laboratory,\\
Chilton, Didcot, Oxon OX11 0QX, U.K.}

\affiliation[e]{Faculty of Nuclear Sciences and Physical Engineering, Czech Technical University,\\
  Prague, 160 00, Czech Republic}
  
  \affiliation[f]{Department of Physics and Astronomy, Uppsala University,\\
  Uppsala, Sweden}

\emailAdd{amit.c@srmap.edu.in}
\emailAdd{sd@ecs.soton.ac.uk}
\emailAdd{ henry.day-hall@cern.ch}
\emailAdd{b.ford@soton.ac.uk}
\emailAdd{s.jain@soton.ac.uk}
\emailAdd{stefano@phys.soton.ac.uk}
\emailAdd{stefano.moretti@physics.uu.se}
\emailAdd{emmanuel.olaiya@stfc.ac.uk}
\emailAdd{claire.shepherd@stfc.ac.uk}

\abstract{The ultimate motivation of our study is to look for signs of physics beyond the Standard Model (BSM). We investigate whether different jet clustering techniques might be more or less suited to the particular final states of interest. In particular, we are interested in fully hadronic final states emerging from the decay chain of the Standard Model like Higgs boson into pairs of light Higgs states, the latter in turn decaying into $b \bar{b}$ pairs. We show that, the ability of selecting the multi-jet final state and to reconstruct invariant masses of the Higgs bosons from it depend strongly on the choice of acceptance cuts, resolution parameters and  reconstruction algorithm as well as its settings. Hence, we indicate the optimal choice of the latter for the purpose of establishing such a benchmark as a BSM signal. We then repeat the exercise for a heavy Higgs boson cascading into two SM-like Higgs states, obtaining similar results.}

\FullConference{%
  41st International Conference on High Energy physics - ICHEP2022\\
  6-13 July, 2022\\
  Bologna, Italy
}

\begin{document}
\maketitle
\section{Introduction}
One of the well known peculiarities of QCD is colour confinement. The quarks and gluons that may appear in the final states in the detector are therefore not in their pure form, and instead forms sprays of colourless hadrons. We therefore, need some clever machinery to provide a map between the set of hadrons in our detector to a set of well defined object called "jets".  A jet definition provides this mapping between hard interaction in our QFT, which is what we are ultimately looking to test. 

The aim of this study is to determine which jet reconstruction tools might be better suited to the final states at the Large Hadron Collider (LHC) in order to extract proper physics. In particular we test different jet clustering techniques for 4$b$ final state coming from the 2-Higgs Doublet Model (2HDM). Furthermore, the 4$b$ final state that we are looking at is an ubiquitous signal of BSM giving access to key features of the underlying BSM scenario, e.g., in the form of the shape of the Higgs potential, hence, of the vacuum stability and perturbative phases of it.

The layout of the paper is as follows. In the next two sections, we discuss different jet reconstruction techniques used, $b$-tagging, the tools used for our simulations and the cutflow adopted. In the one following, we present our results for both signal and background. Then, we conclude.
\section{Jet Clustering Algorithm}
There is a long history associated with these algorithms and first of many jet clustering algorithms was developed in 1977 by Sterman and Weinberg, initially deployed in the context of $e^+ e^- \rightarrow $ hadron scatterings. The type of algorithms currently utilised at the LHC and of particular interest for this study are known as sequential recombination algorithms. Sequential clustering algorithms reduce the complexity of final states by attempting to rewind the showering/hadronisation process.  The algorithms currently used at the LHC are the anti-$k_T$ one and the Cambridge/Aachen (CA) one. 

The more recent variation of the standard jet clustering algorithms is the so-called Variable-$R$ \cite{b}. Variable-$R$ alters the above scheme so as to adapt to events with jets of varying cone size. A modification to the distance measure is made, such that:
\begin{equation}
d_{\rm ij}=min(p_{\rm Ti}^{\rm 2a},p_{\rm Tj}^{\rm 2a})\Delta R^2_{\rm ij} \hspace{0.7cm} d_{\rm iB}=p_{\rm Ti}^{\rm 2a} R_{\rm eff}(p_{\rm T})^2
 \end{equation}
where $R_{\rm eff}(p_{\rm T})=\frac{\rho}{p_T}$ and  $\rho$ is a dimensionful input parameter. There are other two parameters such as $R_{max/min}$, which are cut offs for the maximum and minimum allowed $R_{\rm eff}$. These parameters can be optimised for better results. 

The purpose of introducing variable-$R$ is that in multijet final state where we expect signal $b$-jets with a wide spread of different $p_T$ ’s, we hypothesise that a variable-$R$ reconstruction procedure could improve
upon the performance of traditional fixed-$R$ routines. In particular, using variable-$R$ helps to not rely on a single fixed cone size and suitably engulfs all of the radiation inside a jet, without sweeping up too much outside ‘junk’.

\section{Suitable benchmark, Simulation Details and Cutflow}
We first select a suitable set of parameters in the 2HDM Type-II framework for our model. We consider two sample Benchmark Points (BPs) that we call BP1 and BP2. We work in a scenario where $m_H = 700$ GeV, $m_h = 125$ GeV for BP1 and $m_H = 125$ GeV, $m_h = 60$ GeV for BP2 as we require $m_h < \frac{m_H}{2}$ for $H \rightarrow hh$. Description of the procedure used and cutflow applied for this model are:
 \begin{itemize}
    \item Generate samples of signal events of $\mathcal{O}(10^5)$ for the process 
$gg\rightarrow H \rightarrow hh\rightarrow b\bar{b} b\bar{b}$ using MadGraph5 \cite{c}.
\item Shower and hadronise parton level events using Pythia8 \cite{d}.
\item Apply detector simulation via Delphes CMS card \cite{e}.
\item Perform jet reconstruction, apply cuts on the eflow objects and carry out analysis using MadAnalysis5 \cite{f}.
\item Remove jets with $p_T <$ 50 (BP1) / 20 (BP2) GeV.
\item Implement a simplified MC informed b-tagger on clustered jets after cuts have been applied.
\item Where at least three $b$-jets remain, find the pair best constructing $m_h$ and save as dijet.
\item If four $b$-jets are found, save the remaining pair as a second dijet.
\end{itemize}
\section{Results}
In this section we present detector level results,  we also discuss dominant backgrounds:  QCD 4$b$ production, $gg, q \bar q \rightarrow  Z \bar b b$ and $gg, q\bar q \rightarrow t\bar t$ for signal to background ratio.

As a starting point, we investigate the $b$-jet multiplicity plots for fixed-$R=0.4$ and variable-$R$ with Anti-$k_t$ clustering algorithm for both benchmarks.
\begin{figure}[htb!]
\centering
    \includegraphics[scale=0.4]{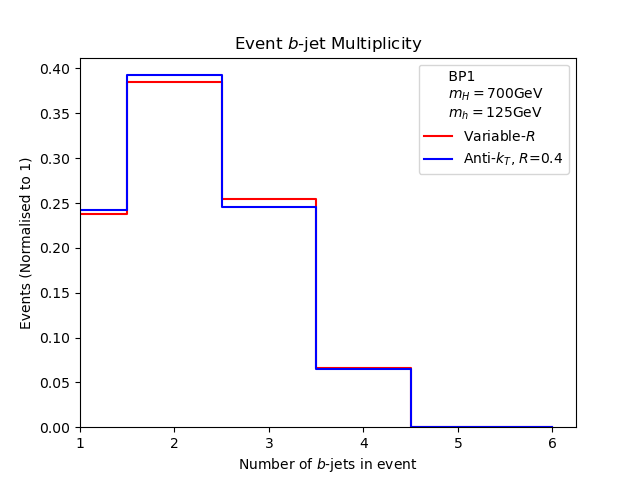}
    \includegraphics[scale=0.4]{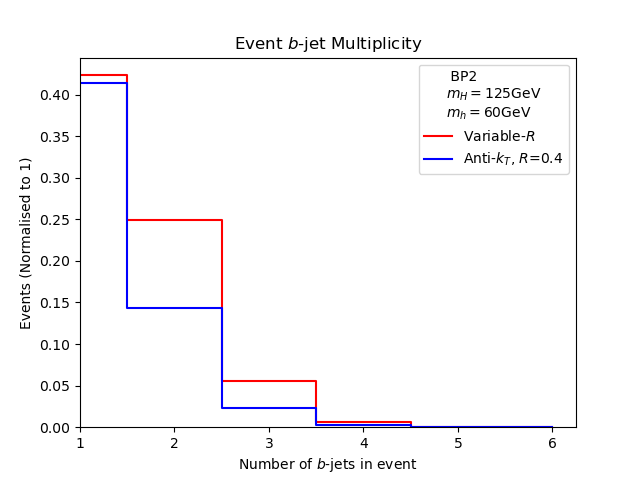}
    \caption{Left panel: The $b$-jet multiplicity distributions for BP1. Right panel: For BP2.}
    \label{fig:pt_30}
\end{figure}
This striking difference between the two plots in Fig.~\ref{fig:pt_30} is due to the relative kinematics of the final state $b$-jets. Due to the different mass configurations, $b$-jets from BP1 have significantly higher $p_T$ than those from BP2, this leads to less events being lost to the trigger as well as from the $p_T$ dependent $b$-tagging efficiencies. It is evident that for variable-$R$, there is a small increase in events reconstructed with higher $b$-jet multiplicity for BP1 while a more significant shift is evident for BP2.

Next, we look at the invariant mass of pairs of clustered $b$-jets (dijets) and fourjets, in order to reconstruct the masses of the resonance from which they originated.

\begin{figure}[htb!]
    \includegraphics[scale=0.45]{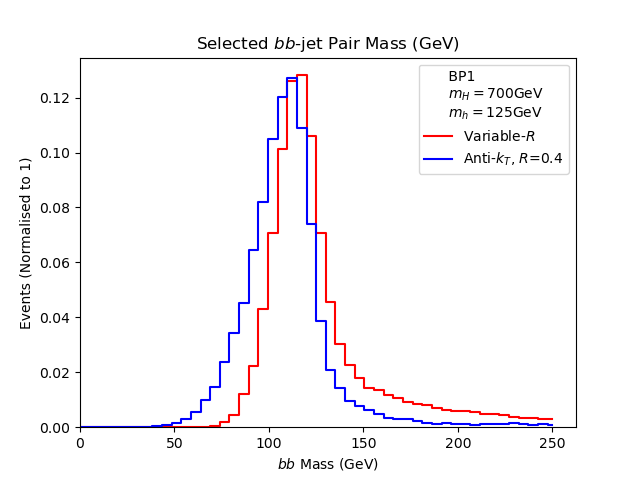}
    \includegraphics[scale=0.45]{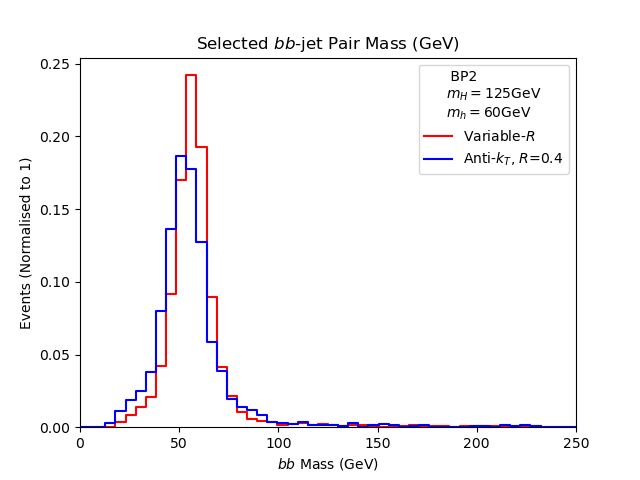}
\\
\includegraphics[scale=0.45]{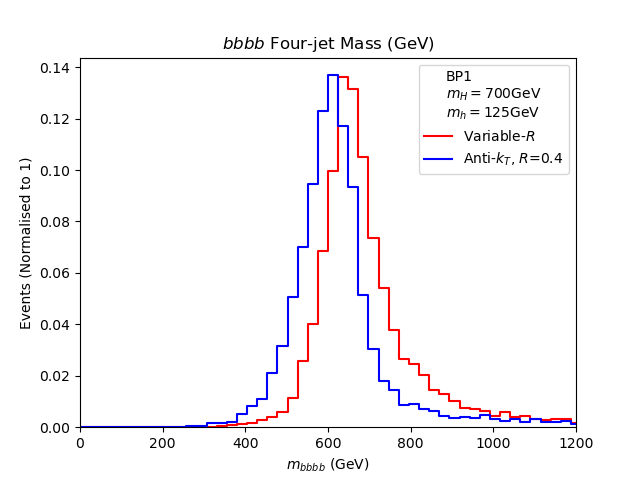}
 \includegraphics[scale=0.45]{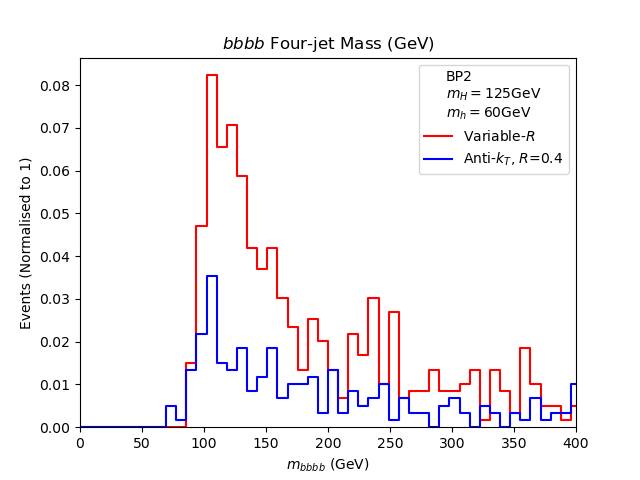}
 \caption{The dijet and four $b$-jet invariant mass distributions for BP1 (left panel) and BP2 (right panel). The peak of the mass distribution obtained from the variable-$R$ algorithm is closer to the MC truth value of the corresponding Higgs.}
 \label{fig:pt_20}
\end{figure}

The dijet and four $b$-jet invariant mass $m_H$ distributions from jet clustering for BP1 (left panel) and BP2 (upper panel) are shown in Fig.~\ref{fig:pt_20}. The peak of the mass distribution obtained from the variable-$R$ algorithm is closer to the MC truth value of the corresponding Higgs resonances and we can see more definitively the benefits of using a variable-$R$ jet clustering algorithm.
\subsection{Signal-to-Background Analysis}
As a final exercise, we perform a calculation of the signal-to-background rates to compare the various jet reconstruction procedures used in our study. To calculate the rates, we generate and analyse $pp \rightarrow b \bar{b} b \bar{b}$, $pp \rightarrow t \bar{t}$ and $pp \rightarrow Z b \bar{b}$ background processes and apply a selection procedure described in Fig.~\ref{fig:pass}. 
\vspace*{1em}
\tikzstyle{node} = [rectangle, rounded corners, minimum width=3cm, minimum height=1cm,text centered, draw=black]
\tikzstyle{arrow} = [thick,->,>=stealth]
\begin{figure}[htb!]
\centering
   \begin{tikzpicture}[node distance=1.5cm]
    \node (n1) [node] {\shortstack{Select events that contain \\exactly four b-jets}}; 
 \node (n2) [node, below of=n1] {Remove event if $|m_{bbbb}- m_H| > 50 $ GeV};
 \node (n3) [node, below of=n2] {\shortstack{Use di-jet pairings chosen in above analysis}};
 \node (n4) [node, below=0.4 cm of n3] {\shortstack{Remove event if $|m_{bb}- m_h| > 20 $ GeV}};
 
     \draw [arrow] (n1) -- (n2);
    \draw [arrow] (n2) -- (n3);
    \draw [arrow] (n3) -- (n4);
   \end{tikzpicture}
   \caption{Event selection used to compute the signal-to-background rates.}
\label{fig:pass}
\end{figure}
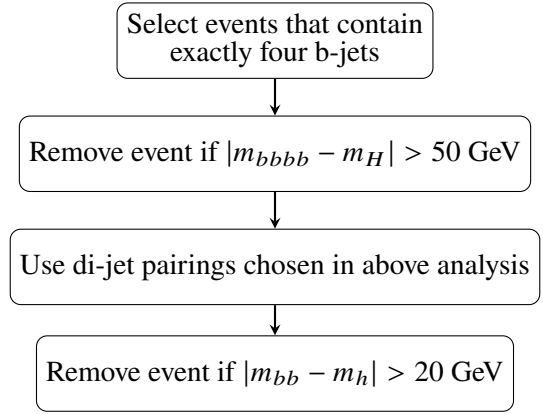

We calculate the significance rates ($\Sigma$) for two values of (integrated) luminosity, given in Table.~\ref{tab:signalbackground5}.
   \begin{table}[!h]
\begin{center}
\scalebox{1}{
\begin{tabular}{ |c|c|c|c| }
 \hline
 & variable-$R$  & $R=0.4$ \\
 \hline
BP1 & 1.881  & 1.366 \\
 \hline
BP2  & 3.707& 1.984   \\
\hline
\end{tabular}
}
\\
\scalebox{1}{
\begin{tabular}{ |c|c|c|c| }
 \hline
 & variable-$R$ & $R=0.4$     \\
 \hline
BP1 &2.753  & 2.000   \\
 \hline
BP2 & 5.426& 2.905   \\
\hline
\end{tabular}
}
\caption{\label{tab:signalbackground5} Upper panel: Final $\Sigma$ values calculated for signal and backgrounds for ${\cal L}=140$ fb$^{-1}$. Lower panel: Final $\Sigma$ values calculated for signal and backgrounds for ${\cal L}=300$ fb$^{-1}$.}
\end{center}
\end{table}
It is then clear from the tables that the variable-$R$ approach works better than fixed-$R$ in improving significance ratios, no matter the choices of R for the latter.

\subsection{Pile-Up}
It has been notable that with variable-$R$  combined with our reduced $p_T$ cuts allows wider cone signal $b$-jets. We therefore perform an analysis by adding Pile-Up (PU) and Multiple Parton Interactions (MPIs) to our event samples.
\begin{figure}[htb!]
	\begin{center}
	\includegraphics[scale=0.4]{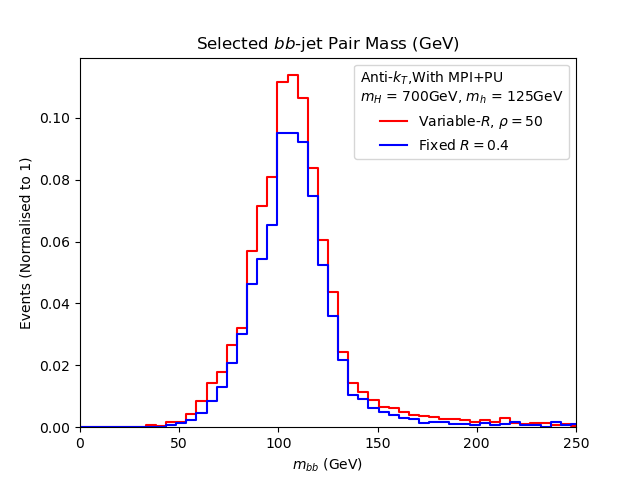}
	\includegraphics[scale=0.4]{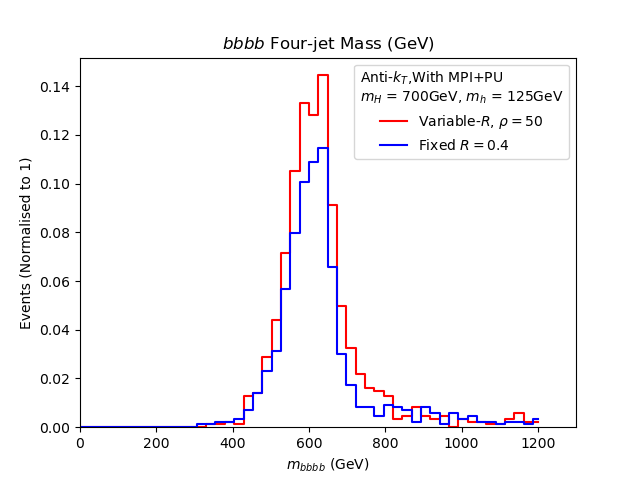}
	 \\
	\includegraphics[scale=0.4]{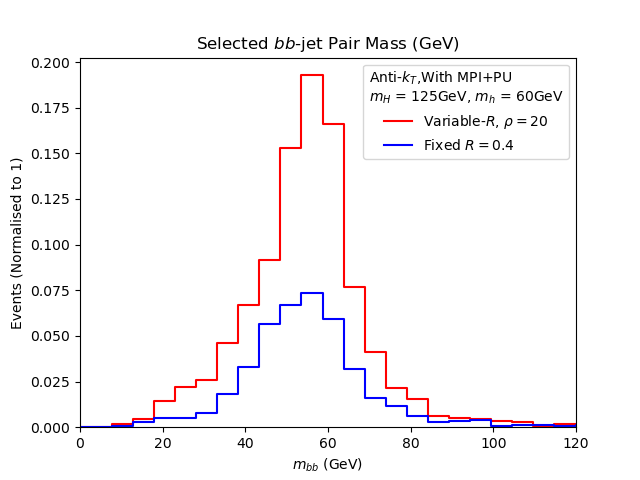}
	\includegraphics[scale=0.4]{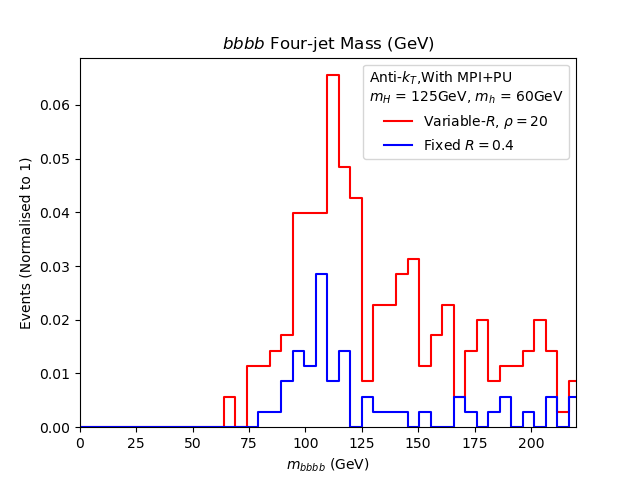}
	 \end{center}
	\caption{Left panel: The $b$-dijet invariant masses for BP1 and BP2, using variable-$R$ and fixed-$R$ clustering, when considering the effect of PU and MPIs. Right panel: The same for the $4b$-jet invariant mass.}
\label{fig:PU_bp2}
\end{figure}

We present in Fig.\ref{fig:PU_bp2} the results for fixed-$R=0.4$ and variable-$R$ jet clustering for completeness. We see that even with PU added, many more events are selected for variable-$R$ jet reconstruction resulting in a better mass peak reconstruction when compared to fixed-$R$. As a final point, we note that a further PU mitigation technique can be done to get rid of extra "junk" which is true for any other jet algorithm.
\section{Acknowledgments}
SM is supported also in part through the NExT Institute and the STFC Consolidated Grant No. ST/L000296/1. BF is funded by the DISCnet SEPnet scholarship scheme. The work of AC is funded by the Department of Science and Technology, Government of India, under Grant No. IFA18-PH 224 (INSPIRE Faculty Award). We all thank G.P. Salam for useful advice. We also give thanks to Benjamin Fuks, Eric Conte and others in the MadAnalysis5 team for their assistance with technical queries. BF and SJ
acknowledge the use of the IRIDIS High Performance Computing Facility, and associated support services at the University of Southampton, in the completion of this work.


\begin{thebibliography}{99}
\bibitem{a}
A. Chakraborty, S. Dasmahapatra, H.A. Day-Hall, B. Ford, S. Jain, S. Moretti, E. Olaiya, C.H. Shepherd-Themistocleous, 
 [arXiv:2008.02499 [hep-ph]] (2020).
 \bibitem{b}
D.Krohn, J. Thaler and Lian-Tao Wang, [	arXiv:0903.0392 [hep-ph]] (2009).
 \bibitem{c}
J. Alwall et al., JHEP 1407 (2014) 079 [arXiv:1405.0301 [hep-ph]].
\bibitem{d}
T. Sjostrand, S. Mrenna and P. Z. Skands, Comput. Phys. Commun. 178 (2008)
852 [arXiv:0710.3820 [hep-ph]].
\bibitem{e}
J. de Favereau et al., [arXiv:1307.6346 [hep-ex]] (2013).
\bibitem{f}
 E. Conte and B. Fuks, Int. J. Mod. Phys. A 33 no. 28 (2018) 1830027
[arXiv:1808.00480 [hep-ph]].
\end{thebibliography}
\end{document}